\documentstyle[twocolumn,aps,prl,epsf,floats,psfig,epsfig]{revtex}

\begin{document}

\wideabs{

\title{The Cerium volume collapse: Results from the LDA+DMFT approach}

\author{K. Held,$^1$ A.K. McMahan,$^2$ and R.T. Scalettar$^3$}
\address{$^1$Physics Department, Princeton University, Princeton, NJ 08544}
\address{$^2$Lawrence Livermore National Laboratory, University of California,
Livermore, CA 94550}
\address{$^3$Physics Department, University of California, Davis, CA 95616}
\date{\today}

\maketitle
\begin{abstract}
The merger of density-functional theory in the local density
approximation (LDA) and many-body dynamical mean field theory (DMFT)
allows for an  {\em ab initio} calculation of Ce including the inherent
$4f$ electronic correlations.  We solve the  DMFT equations
 by the quantum Monte Carlo (QMC) technique and calculate the Ce
energy, spectrum, and double occupancy as a function of volume.  At low
temperatures, the correlation energy exhibits an anomalous region of
negative curvature which drives the system towards a thermodynamic
instability, i.e., the $\gamma$-to-$\alpha$ volume collapse, consistent
with  experiment. The connection of the energetic with the spectral
evolution shows that the physical origin of the energy anomaly and, thus,
the volume collapse is the
appearance of a quasiparticle resonance in the
$4f$-spectrum which is accompanied by a rapid growth in the double
occupancy.\\

\end{abstract}


}


Cerium exhibits the well known $\gamma$-$\alpha$ phase transition
characterized by an unusually large volume change of 15\% \cite{EXPT}.
Similar volume collapse transitions are observed under pressure in Pr
and Gd (for a recent review see \cite{JCAMD}).  It is widely believed
that these transitions arise from changes in the degree of $4f$
electron correlation, as is reflected in both the Kondo volume
collapse\cite{KVC} and the Mott transition\cite{JOHANSSON} models.
The former ascribes the collapse to a rapid change in
the valence-electron screening of the local $4f$-moment, which is
accompanied by the appearance of an Abrikosov-Suhl-like quasiparticle
peak at the Fermi level, lying between the remaining Hubbard-split $4f$
spectral density.  In other words, the Kondo temperature of the
$\alpha$-phase is much larger than  that of the  $\gamma$-phase.  While
originally formulated in terms of the Anderson impurity model, similar
rapid thermodynamic and spectral changes are seen for the lattice
version, or periodic Anderson model\cite{HUSCROFT}.

The Mott transition model envisions a more abrupt change from
itinerant, bonding character of the $4f$-electrons in the
$\alpha$-phase to non-bonding, localized character in the
$\gamma$-phase, driven by changes in the $4f$-$4f$ inter-site
hybridization.  Thus, as the ratio of the $4f$ Coulomb interaction to
the $4f$-bandwidth increases with increasing volume, a Mott transition
occurs to the $\gamma$-phase.  While originally motivated by the
Hubbard model, most recent support for this perspective has come from
orbitally polarized\cite{ORBPOL} and self-interaction
corrected\cite{SIC} modifications of local-density functional theory.
It has been argued\cite{JCAMD} that these modified local-density
methods resemble static mean field treatments in which the 
$\gamma$-phase is spin and orbitally polarized such that the 14 $4f$-bands are
Hubbard split into one band below and 13 above the Fermi level for Ce.
At smaller volumes, the mean-field polarization disappears and, thus,
the $\alpha$-phase resembles the ordinary LDA solution with all $14$
bands grouped together just above but slightly overlapping the Fermi
level.  While this approach gives the correct energy in the limit of
small volume, and also at low temperature for large volume (though not
the correct paramagnetic phase), a true many-body solution would allow
for a central quasiparticle peak in the presence of the Hubbard
splitting as observed in the Mott transition of the one-band Hubbard
model \cite{DMFT2,GEBHARD}.  The Kondo volume collapse calculations
\cite{KVC}, on the other hand, take such electronic correlations into
account but are based on simplified models.

In this situation, the  recently developed merger \cite{LDADMFT} of LDA
\cite{JG} and DMFT \cite{DMFT1,DMFT2} offers an ideal means
to study Ce realistically, including the critical intra-site $4f$
electron correlations \cite{INTERSITE}.  Only two very recent LDA+DMFT
calculations have been reported to date for $f$-electron systems.
Savrasov, Kotliar, and Abrahams \cite{SAVRASOV} have employed an
interpolation scheme to the DMFT self-energy inspired by the iterative
perturbation theory (IPT), and present  total energy calculations
 and the spectrum              for Pu.
Z\"olfl {\it et al.} \cite{ZOLFL} have used the non-crossing
approximation (NCA) to report the first Ce $\alpha$- and 
$\gamma$-phase spectra.
Both papers find
Hubbard splitting, with an additional  quasiparticle peak at
the Fermi level, for the respective low-volume $\alpha$-phases of the
two materials.   
For LaTiO$_3$, a metal close to a Mott transition, LDA+DMFT
calculations employing the IPT and NCA approximations to solve the
DMFT equation have been compared to the more rigorous
QMC\cite{METHOD} treatment
and quantitative differences have been reported \cite{NEKRASOV}.

In order to
analyze the Ce volume collapse,
the present paper  reports the first LDA+DMFT(QMC) calculations of the Ce total
      energy, investigating a wide range of volume and temperature. We  find
the low-temperature total energy to exhibit a distinctive feature which is
consistent with the observed Ce volume collapse.
Additional calculations  of the spectral function,
imaginary time Green's function, and double occupancy show
that the energy feature coincides
with the rapid growth of both the quasiparticle peak and the double
occupancy.    All these signatures moderate with increasing
temperature. To
provide insight into the modified local-density methods, we also
contrast these LDA+DMFT(QMC) results with Hartree-Fock (HF) or static mean
field solutions of the same Hamiltonians.

The DMFT(QMC) calculations in this paper solve the Hamiltonians
\begin{eqnarray}
H = &&\sum_{{\bf k},lm,l' m',\sigma} (H^0_{\rm LDA}({\bf k}))_{lm,l' m'}
\,\hat{c}^\dagger_{{\bf k}\,lm\sigma} \hat{c}^{ }_{{\bf k}\,l' m'\sigma}
\nonumber \\
&&+\; \frac12 \, U_f \!\!\! \sum_{i,m\sigma,m'\sigma'} \!\!\!\!\!\!\! ^{'} \,
\hat{n}_{ifm\sigma}\, \hat{n}_{ifm'\sigma'} \, ,
\label{Ham}
\end{eqnarray}
where {\bf k} are Brillouin zone vectors, $i$ are lattice sites, $lm$
denote the angular momentum, $\sigma$ is the spin quantum number,
$\hat{n}_{ifm\sigma} \equiv \hat{c}^\dagger_{ifm\sigma} \hat{c}^{
}_{ifm\sigma}$, and the prime signifies $m\sigma \neq m' \sigma'$.  The
$16\times16$ matrices ${H}^0_{\rm LDA}({\bf k})$ denote the matrix
elements of the LDA Hamiltonian w.r.t. the 16 orbitals $(6s, 6p, 5d,
4f)$ for fcc Ce, as described in Sec.~4.2 of Ref.~\onlinecite{JCAMD}.
The $4f$ site energies are shifted to avoid double counting of the
$4f$-$4f$ Coulomb interaction which is explicitly incorporated in
Eq.(\ref{Ham}) via the $U_f$ term. These shifted site energies
 and the
screened $4f$-$4f$ Coulomb interactions $U_f$ were obtained by
companion constrained-occupation calculations, and their values
together with the effective  $4f$ electron bandwidth are shown in
Fig.~5 of Ref.~\onlinecite{JCAMD} as a function of volume.  Since the
$4f$-orbitals are well localized, uncertainties in $U_f$ and the $4f$
site energies are relatively small, and, given the significant
volume-dependence of the $4f$ site energy, only translate into a possible
volume shift.  
We have not included the spin-orbit interaction which
has a rather small impact on LDA results for Ce, nor the intra-atomic
exchange interaction which is less relevant for Ce as occupations with
more than one $4f$-electron on the same site are rare.

For the present paramagnetic calculations we take the $4f$ self-energy
matrix to be diagonal $\Sigma(i\omega)\, \delta_{m\sigma,m'\sigma'}$
and use two complementary approaches to perform
the transformation from $G(\tau)$ to $G(i\omega_n)$. The approach described in \cite{ULMKE} is used
on a volume subgrid to validate a new faster approach
which fits the $G(\tau)$ data with basis functions 
of the form $e^{-\tau\epsilon_i}/(e^{-\epsilon_i/T}+1)$
and which is employed for the full
temperature and volume grid.
Unless noted otherwise, our results are all 
extrapolated to the limit of zero
imaginary time discretization  $\Delta \tau \rightarrow 0$
in the QMC.
The  DMFT energy per site was evaluated from
\begin{eqnarray}
E_{\rm DMFT}&\! =\!& \frac{T}{N} \sum_{n {\bf k} \sigma}
{\rm Tr}({ H}^0_{\rm LDA}({\bf k}) { G}_{\bf k}(i\omega_n))
e^{i\omega_n 0^+} + U_f \, d.
\label{Eng}
\end{eqnarray}
Here, Tr denotes the trace over the $16\times16$ matrices, $T$ the
temperature, $N$ the number of ${\bf k}$ points, and 
\begin{equation}
d = \frac12
{\sum}_{m\sigma,m'\sigma'}' \langle \hat{n}_{ifm\sigma}\,
\hat{n}_{ifm'\sigma'}\rangle
\label{double}
\end{equation}
is a generalization of the one-band
double occupation for multi-band models, calculated directly in the
QMC and 
related to the local magnetic moment via $\langle  {m_Z}^2 \rangle =
\sum_{m\sigma} \langle\hat{n}_{if m\sigma}\rangle - (2/13)\, d$.

\begin{figure}[htb]
 \epsfxsize=1.13\hsize \epsfbox{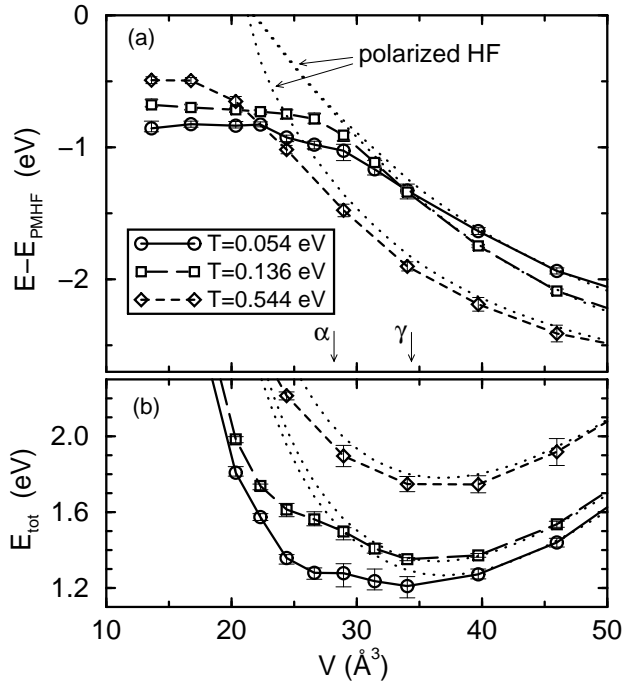}

\caption{(a) Correlation energy $E_{\rm DMFT}\!-\!E_{\rm PMHF}$
as a function of atomic volume (symbols) and 
polarized HF
energy $E_{\rm AFHF}\!-\!E_{\rm PMHF}$ (dotted lines which, at large V, approach the
DMFT curves for the respective temperatures);
arrows:  observed volume collapse from the $\alpha$- to the $\gamma$-phase.  
The correlation energy sharply bends away from the
polarized HF energy in the region of the transition. (b) The resultant negative curvature leads to a growing depression
     of the total energy near $V\!=\!26$--$28$ \AA$^3$ as temperature
     is decreased, consistent with an emerging double well at still
     lower tempertures and thus the $\alpha$-$\gamma$ transition. The
     curves at $T=0.544\,$eV were shifted downwards by $-0.5\,$eV to match the energy range.
\label{figE}}
\end{figure}

Fig.~\ref{figE}a shows our calculated DMFT(QMC) energies $E_{\rm DMFT}$
as a function of atomic volume at three temperatures  {\it relative}
to the paramagnetic HF energies $E_{\rm PMHF}$ of  Eq.(\ref{Ham}), 
i.e., the energy contribution
due to electronic correlations. 
Similarly given are the polarized HF energies
which reproduce $E_{\rm DMFT}$
at large volumes and low temperatures.
With decreasing  volume, however,  the DMFT
energies bend away from the polarized HF solutions.
This striking effect becomes more pronounced, and begins at slightly
larger volume, as temperature is decreased.  At $T\!=\!0.054\,$eV, 
a region of negative curvature in $E_{\rm
DMFT}\!-\!E_{\rm PMHF}$ is evident within the observed two phase region (arrows).

Fig.~\ref{figE}b presents the calculated LDA+DMFT total energy $E_{\rm
tot}(T)\!=\!E_{\rm LDA}(T)\!+\!E_{\rm DMFT}(T)\!-\!E_{\rm mLDA}(T)$
where $E_{\rm mLDA}$ is the energy of an LDA-like solution of the model
Hamiltonian in Eq.(\ref{Ham}) \cite{mLDA}.  Since both $E_{\rm LDA}$
and $E_{\rm PMHF}\!-\!E_{\rm mLDA}$ have positive curvature throughout
the volume range considered, it is the negative curvature of the
correlation energy in Fig.~\ref{figE}a which leads to the dramatic
depression of the LDA+DMFT total energies in the range $V\!=\!26$--$28$
\AA$^3$ for decreasing temperature, which contrasts to the smaller
changes near $V\!=\!34$ \AA$^3$ in Fig.~\ref{figE}b.  This trend is
consistent with a double well structure emerging at still lower
temperatures (prohibitively expensive for QMC simulations), 
and with it the concomitant volume collapse transition.
The general shallowness of our $T\!=\!0.054\,$eV$\!=\!632\,$K isotherm
in Fig.~\ref{figE}b is consistent with the $\sim$550$\,$K critical end
point for the $\alpha$-$\gamma$ transition, as is the width of this
region compared to the indicated room temperature volumes of the
$\alpha$ and $\gamma$ phases.  We estimate entropy corrections $TS$
would alter the shape of this curve by less that 0.1 eV.  Furthermore,
pressure-volume results (not shown) obtained from smoothed fits to our
$632$-K energies yield 70-110\% of the observed 9.6 \AA$^3$ 
change in atomic volume between $\gamma$- and $\alpha$-phases at 0 and 5 GPa,
respectively, and they track the experimental isotherm to within 2 and
in some cases 1 \AA$^3$ from 5 to 50 GPa. Altogether, this is
reasonable agreement with experiment given our use of energies rather
than free energies, the different temperatures, and the  LDA and DMFT
approximations.

\begin{figure}[htb]
\begin{center}
\hspace{-1.cm} \epsfxsize=1.\hsize \epsfbox{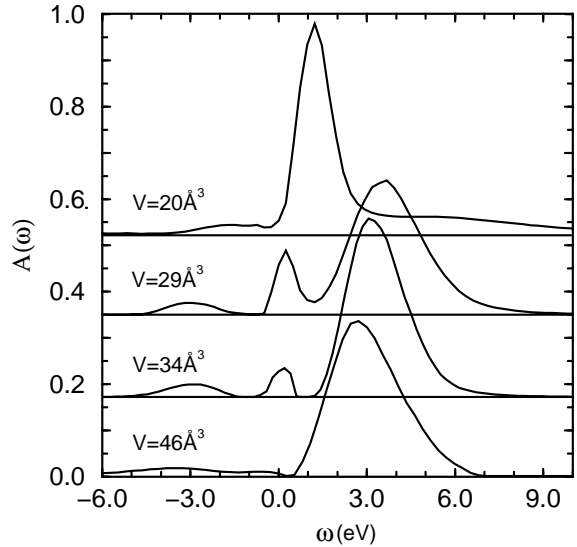}
\end{center}
 \vspace{-0.6cm}
\caption{
Evolution of the $4f$ spectral function $A(\omega)$ with
volume at $T\!=\!0.136\!$ eV ($\omega\!=\!0$ corresponds to the chemical potential;
curves are offset as indicated; $\Delta
\tau\!=\!0.11\!$ eV$^{-1}$). Coinciding with the sharp anomaly in the correlation
energy (Fig.~\ref{figE}), the central quasiparticle resonance   disappears.
\label{figSpec}}
\end{figure}

To clarify the physical origin of the sharp energy feature, we employ
the maximum entropy method to study the evolution of the $4f$ spectral
function $A(\omega)$ with atomic volume.  At $V\!=\!20$ \AA$^3$,
Fig.~\ref{figSpec} shows that almost the entire spectral weight lies in
a large quasiparticle peak with a center of gravity slightly above the
chemical potential.  This is similar to the LDA solution, however, a
weak upper Hubbard band is also present even at this small volume.  At
the volumes $29$ \AA$^3$ and $34$ \AA$^3$ which approximately bracket
the $\alpha$-$\gamma$ transition, the spectrum has a three peak
structure which consists of a quasiparticle peak or Abriksov-Suhl
resonance at the Fermi energy, in addition to the two Hubbard side
bands at about $\pm 3\,$eV.  The quasiparticle peak is seen to
dramatically shrink in going from $V=29$ \AA$^3$ to $V=34$ \AA$^3$,
which coincides with the range of negative curvature in the correlation
energy.  Finally, by $V=46$ \AA$^3$, the central peak has disappeared
leaving only the lower and upper Hubbard bands in the spectrum.  

\begin{figure}[htb]
 \vspace{-1.0cm}
 \hspace{-1cm} \epsfxsize=1.1\hsize \epsfbox{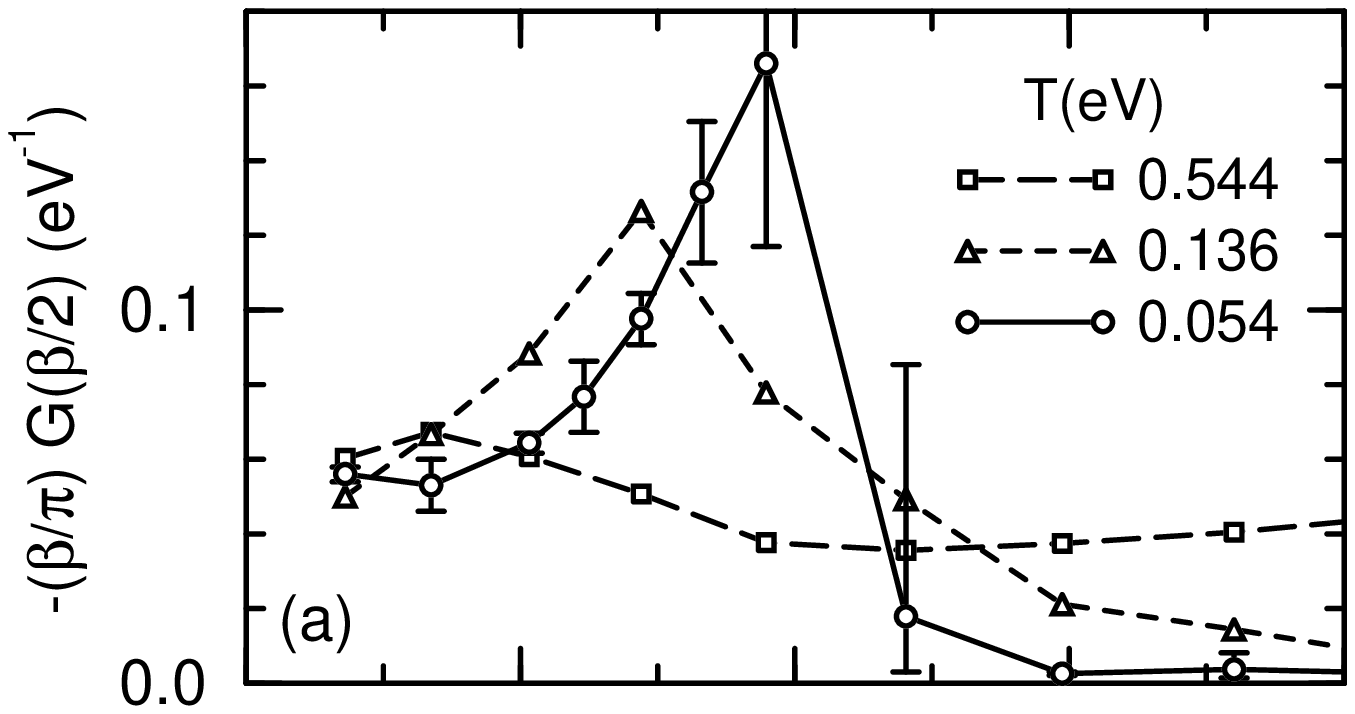}
 \vspace{-4.84cm}

  \hspace{-1cm} \epsfxsize=1.1\hsize \epsfbox{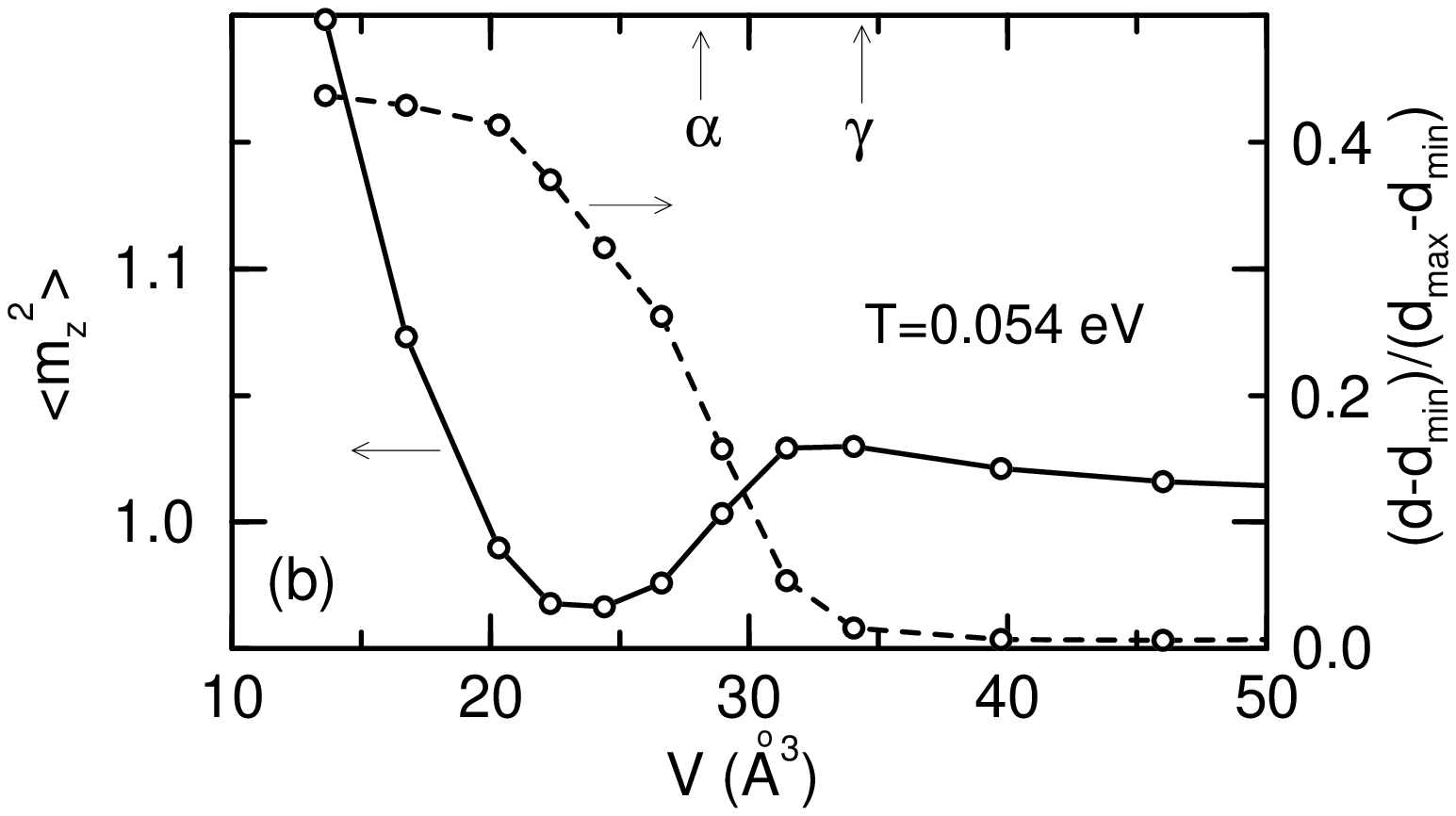}
 \vspace{-2.3cm}
\caption{(a) $-(\beta/\pi)\,G(\beta/2)$ vs.~volume, which at low
temperature gives the $4f$ spectral weight at the chemical potential;
(b) 
Local magnetic moment and ratio of the double occupancy vs.~volume.
The latter indicates that the $\gamma$-$\alpha$ transition coincides
   with the onset of delocalization.}
\label{figD}
\end{figure}

An alternative quantity that allows the study of the spectrum close to the
Fermi energy is the Green function at imaginary time
$\tau\!=\!\beta/2$, since in the limit of low temperature the density
of states at the Fermi level is given by
$N(0)=-(\beta/\pi)\,G(\beta/2)$ \cite{TRIVEDI}.  With increasing
volume, the low-temperature results in Fig.~\ref{figD}a show first an
increase in the quasiparticle peak  around the chemical potential
due to narrowing of the LDA $4f$-bands, followed  by a sharp drop at the $\alpha$-$\gamma$ transition, supporting the
results of Fig.~\ref{figSpec}.

To measure the itinerant or localized 
character we study the generalized double occupancy $d$ of
Eq.~(\ref{double}). For fully localized spins,
$d$ takes its smallest value at
   the minimal fraction of sites having two $4f$-electrons which
is $d_{min}\!=\!\max(0,n_f\!-\!1)$ 
for $n_f\leq 2$.
In contrast, for a fully itinerant system in the uncorrelated
   $U_f$=0 limit, the maximal value $d_{max}\!=\!(13/28)n_f^2$ is
   obtained, corresponding to $\langle n_{m\sigma}
   n_{m'\sigma'}\rangle\!=\!(n_f/14)^2$.
With decreasing volume, Fig.~\ref{figD}b clearly shows 
 a dramatic increase in the ratio ($d\!-\!d_{min})/(d_{max}\!-\!d_{min})$
associated to the onset of delocalization. However, the delocalization 
is not completed and $d$ is still considerably lower than its
maximal value, even at the lowest volumes of  Fig.~\ref{figD}b.
This reflects the correlated nature of the
$\alpha$-phase with a reduced Coulomb interaction energy $U_f\, d$
and thus, compared to the uncorrelated static mean field solution,
a lower DMFT energy in  Fig.~\ref{figE}a.
  The rapid increase in double occupancy implies that 
  the local magnetic moment shows a considerable change
  at the transition (the increase for small volumes is due to
  an increase in $n_f$) which is, however, less pronounced than in the
  static mean field theories.

In conclusion, our LDA+DMFT(QMC) calculations for Ce show an anomaly 
in the correlation energy
leading to a shallowness in the total energy close to the critical
endpoint for the $\alpha$-$\gamma$ transition, and suggest an
emerging double well as temperature is further decreased,
consistent with the observed transition.
 The $4f$-spectra and a measure of the
$4f$-spectrum at the Fermi level show Hubbard splitting in the
large-volume $\gamma$-phase, with an Abrikosov-Suhl-like quasiparticle
peak first appearing at the Fermi level in the transition region, and
then growing at the expense of the Hubbard side-bands with subsequent
compression in the $\alpha$-phase.  These are characteristic attributes
of the Kondo volume collapse picture for Ce \cite{KVC}.  On the other
hand, we also find a rapid increase in the double occupancy at low
temperature in going from the $\gamma$- to the $\alpha$-phases, which could
be interpreted as increased itinerancy of the $4f$-electrons, a tenet
of the Mott transition picture \cite{JOHANSSON}.  There may well be
greater similarities between the two scenarios than has been accepted,
as argued recently in a comparison of many-body solutions of the
respective Anderson and Hubbard model paradigms for these pictures
\cite{HELD}.

Our comparison of HF to the DMFT energies also offers insight into
the modified local-density calculations for the Ce transition
\cite{ORBPOL,SIC,LDA+U}, which resemble static mean field treatments.  Our
results suggest that polarized solutions  can give good
low-temperature energies at large volume for the $\gamma$-phase,
however, may be offset in energy in the low-volume paramagnetic
$\alpha$-regime.  Even so, it is tempting to speculate from Fig.~\ref{figE}a
that their slopes are correct, which would reconcile standard LDA or
generalized gradient extensions \cite{SODERLIND} doing so well for the
volume dependence in the itinerant phases, even though they can not
capture the residual Hubbard splitting.  

We acknowledge support by the Alexander von Humboldt foundation (KH)
and NSF-DMR-9985978 (RTS).  Work by AKM was performed under the
auspices of the U.S.~DOE by U.~Cal., LLNL under contract No.
W-7405-Eng-48.  We are grateful for the QMC code of \cite{DMFT2} (App.~D)
 which was modified for use in part of the present
work,  to A.~Sandvik for making available his maximum
entropy code, and to D.~Vollhardt and G.~Esirgen for useful discussions.


\begin{references}
\bibitem{INTERSITE}
For Ce, intersite correlations are less important since the Coulomb
interaction is between well localized $4f$-electrons and  the volume
collapse transitions occur at room temperature well above any magnetic
order.
\bibitem{mLDA}
 We solve self-consistently for 
$n_f$ using a $4f$ self energy $\Sigma =U_f(n_f-\frac12)$, and then 
remove this contribution from the
eigenvalue sum to get the kinetic energy.  The potential energy is
taken to be $\frac12 U_f n_f(n_f-1)$.
\end{references}
\end{document}